\documentclass[submission,copyright,creativecommons]{eptcs}
\providecommand{\event}{IMPEX  2017} 
\usepackage{breakurl}             
\usepackage{underscore}           

\usepackage{bsymb,b2latex}

\usepackage{graphicx,color}

\newenvironment{Bcode}
  {\begin{array}{|@{\quad}l@{\quad}|}\hline\\[-1.8\jot]}
  {\\[1.8\jot]\hline\end{array}}

\title{Explicit Modelling of Physical Measures:\\ 
From Event-B to Java}
\author{J Paul Gibson
\institute{SAMOVAR  UMR 5157\\ T\'el\'ecom Sud Paris, Evry, France}
\email{paul.gibson@telecom-sudparis.eu}
\and
Dominique M\'ery
\institute{LORIA UMR 7503\\ Telecom Nancy,  Universit\'e de Lorraine\\
  Vand\oe uvre-l\`es-Nancy, France}
\email{dominique.mery@loria.fr}
}
\def\titlerunning{Explicit Modelling of Physical Measures}
\def\authorrunning{J. P. Gibson \& D. M\'ery}
\begin{document}
\maketitle

\begin{abstract}

The increasing development of cyber-physical systems (CPSs) requires modellers to 
represent and reason about physical values. 
This paper addresses two major, inter-related, aspects that arise
when modelling physical measures.
 Firstly, there is often a {\it heterogeneity 
 of representation}; for example: speed can be represented in  many different units (mph, kph, mps, etc\ldots).
 Secondly, there is {\it incoherence in composition}; for example: adding a speed to a temperature
 would provide a meaningless result in the physical world, even though such a purely mathematical operation
 is meaningful {\it in the abstract}.
These aspects are problematic
when implicit  semantics --- concerned with measurements --- in CPSs are not
explicit (enough) in the requirements, design and implementation models.
We present an engineering approach for 
explicitly modelling measurements during all  phases
of formal system development. We illustrate this by moving from Event-B models  to 
Java implementations, via object oriented design.

\end{abstract}

\section{Introduction}

\subsection{Problems  when using  implicit measurements and dimensions}

Historically, humans have made many critical mistakes when using units of measurement.
For example, Chrostopher 
Columbus miscalculated the circumference of the earth when he used Roman miles instead of nautical miles~\cite{Rickey92}.
Since software has been written,
there is added risk of bugs arising out of the lack of explicit measurement types (and dimensions) in
programs. Perhaps the best known bug of this type
is the Mars climate orbiter error~\cite{SauserRS09}, where separate software modules
were using different units of measurement (imperial and metric); yet when they communicated data they were not aware of the
different representations and wrongly assumed that the data being shared was represented in the same unit/scale.
This issue would have been avoided if the units of measurement had been explicitly (and formally)
specified in the module interfaces.

\subsection{Proposed requirements  for explicit modelling of measures}
\label{subsec:Requirements}

Formal methods are often used in the development of critical software systems~\cite{BowenStavridou93}; but they
often permit the construction of  models which are mathematically meaningful, in the abstract, but which
have no coherent interpretation  in the real  (concrete) world.
We propose  a modelling technique which helps software engineers to avoid
building models that are incoherent, by  explicitly specifying measurement dimensions and scales (units).

We are motivated by the international standard for physical measurements which identifies seven
 base units/dimensions and numerous derived units and constants~\cite{Quinn95,ThompsonTaylor08}.
 We require that the measurement model must:
{\bf (1)} Specify seven  base units (explicit) in a standard library: mass, length, time, temperature, current, light and matter.
{\bf (2)} Support the derivation of  (standard) units  that can be added explicitly to the library (as required) or used implicitly.
{\bf (3)} Support definition of non-standard units that can be defined in terms of standard units.
{\bf (4)} Facilitate specification of scalar universal constants (like PI).
{\bf (5)} Facilitate  specification of  physical constants of certain dimension (like the speed of light in a vacuum).
{\bf (6)} Be amenable to static/automated analysis for detection of
incoherent mixing of units (like adding a speed and a temperature).
{\bf (7)} Support implicit scaling - being able to add/subtract/compare measurements in the same dimension but different scales, without the need to  explicitly scale them.

We choose to use Event-B~\cite{Abrial10} contexts as our modelling language because 
we wish to re-use our measurement models to play the role of a domain ontology~\cite{GruberOlsen94} in the 
development of safety-critical CPSs using refinement~\cite{AbrialHallerstede07}. 
We also wish to have a powerful  reasoning tool (in this case Rodin~\cite{AbrialBHHMV09}) for
automating validation and verification.
Our modelling approach
could be easily followed using  other formal methods (that provide similar language and tool support).

\section{Related work}
\label{sec:RelatedWork}

The thesis by Kennedy~\cite{Kennedy96} is one of the earliest works to propose a solution
to the problem of representing dimensions in programming languages.
The thesis establishes and proves a formal 
relationship between conventional type systems and the dimension type systems (using a functonal
programming language as its foundations). However, the work does not
 not cover the situation of multiple systems of units within the same dimension ( formal engineering requirements 3 and 7, in section \ref{subsec:Requirements}).

In `{\it Object-Oriented Units of Measurement}'~\cite{AllenCLMS04}, the authors state:
{\it ``Physical units and dimensions are a commonly used computational construct in science and engineering, but there is relatively little support for them in programming languages.''}
They demonstrate how an OO language that offers rich semantics for polymorphism and dynamic binding  can use explicit typing to model a hierachy of measurements
for physical values. Their approach also does not meet requirements  3 and 7.

In `{\it Arithmetic with Measurements on Dynamically-typed Object-oriented Languages}'~\cite{WilkinsonPR05}, the authors
implement - in Smalltalk - a measurement system that is general enough to include the SI units for physical measurements, as well as
non-physical measures such as units of currency (see section \ref{sec:Currency}). 
The choice of Smalltalk is interesting because the language's dynamic type system gives the
programmer more support for defining new types, but restricts the
strength of static analysis that can be done. Their approach does not support our requirements 2,4, and 5.

In `{\it Osprey: A Practical Type System for Validating Dimensional Unit Correctness of C Programs}''\cite{JiangSu06},
the authors present a sound type system to automatically check for all potential errors involving units of measurements.
The system is based on annotating the C code with unit/dimension 
typing constraints that are then checked using a constraint solver. However, their analysis is limited to comparing units and
does not  support comparing dimensions. Their approach only partially meets all our requirements, and introduces 
complexity of annotations into the programming task.

In ``{\it A model-driven approach to automatic conversion of physical units}''~\cite{CooperMcKeever08},
the authors propose checking physical units at the level of the modelling language, removing the need for such a support in the underlying implementation language. They choose to express all physical values in the standard canonical form,
as  products of powers of the SI base units, and present an elaborate algorithm for translating to and from such a form for any
arbitrary unit. However, since in CellML (the modelling language upon which their approach is built)
 all constants and variables are explicitly annotated with their units, they have not needed to make any type inferences.
Thus, they have  not  considered meeting our 2nd requirement.

\section{Formal Specification in Event-B}

In this section we present three different Event-B
 contexts which are used to meet our
measurement modelling requirements, as specified in subsection\ref{subsec:Requirements}.
We chose to structure the model  by separating the representation of real-world values
(in context  {\tt C\_Values})
from the real-world dimensions (in context  {\tt C\_Dimensions}).
The context   {\tt C\_Measurements} then composes them to provide a standard  model
that can then be extended with derived measurements and values.

\subsection{Values as floats}

\begin{figure}[htb]
\centering
\fbox{
\includegraphics[scale=.40]{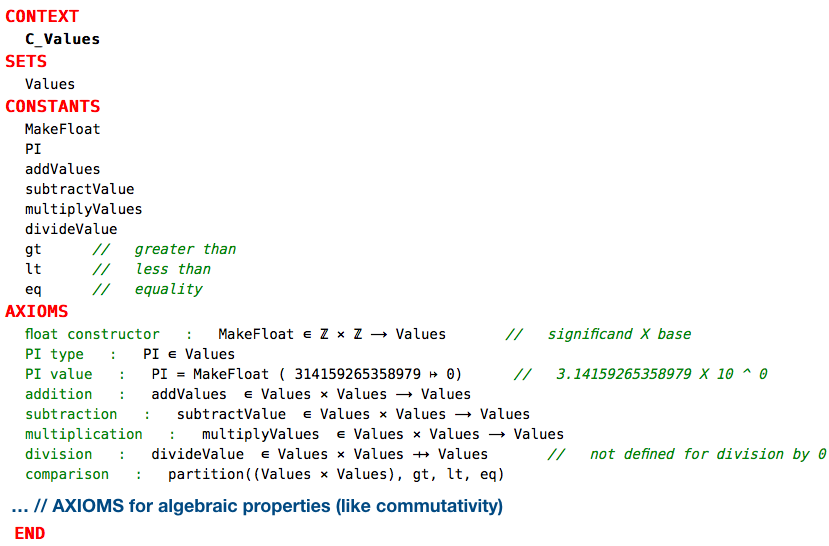}}
\caption{The floating number (partial) abstract model}
\label{fig:Values}  
\end{figure}

We define a context {\tt C\_Values}  to provide a concrete  mechanism for constructing and representing
values as floating point decimal numbers~\cite{Goldberg91}.
A partial specification of the model  is given in Figure~\ref{fig:Values}.The constant {\tt PI} is used as a concrete example of how we represent values using floating point numbers.
The first integer,  commonly know as the
{\it significand}, is to be interpreted as a float with the floating point occurring after the first decimal digit.
The second integer is to be interpreted as the power of 10, commonly known as the {\it base},
 which is to be multiplied to the {\it significand} in order to give the real value of the  floating point number
 ($significand \times 10^{base}$).
The arithmetic operators are defined algebraically (including  axioms - not shown  -
for operator properties, such as commutativity, that
are needed for later proofs). 
 We note that this context makes no reference to physical units or dimensions.

\subsection{Dimensions}

\begin{figure}[htb]
\centering
\fbox{
\includegraphics[scale=.40]{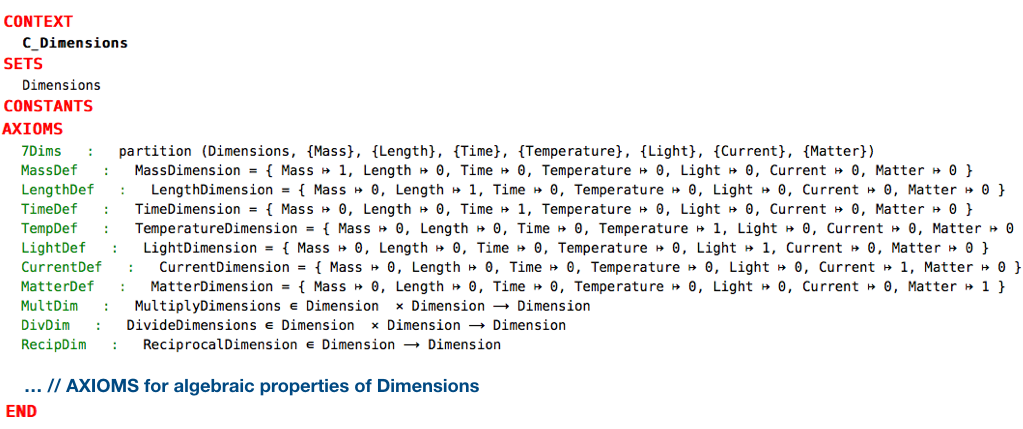}}
\caption{The dimension model -  {\tt CONTEXT C\_Dimensions  AXIOMS}}
\label{fig:Dimensions}  
\end{figure}

The partial specification of the context  {\tt C\_Dimensions} is given  in Figure~\ref{fig:Dimensions}.The seven SI dimension units~\cite{Quinn95,ThompsonTaylor08} - {\tt Mass}, {\tt Length}, {\tt Time}, {\tt Temperature}, {\tt Light},
{\tt Current} and {\tt Matter} are explicitly modelled, and we can  derive new dimensions
through multiplication, division and reciprocation. 
So, for example a derived dimension for {\tt Acceleration}  would be defined as:
Length  divided by Time squared. So the associated powers would be 1 for {\tt Length} and -2 for {\tt Time}, with all other 
base dimensions having 0 powers giving: 
\small
$\{Mass \mapsto 0,    Length \mapsto 1,  Time \mapsto 2, Temperature \mapsto 0, Light \mapsto 0, Current \mapsto 0, Matter \mapsto 0\}$.
\normalsize
We note that this context makes no references to physical values.

\subsection{Measurements}

\begin{figure}[htb]
\centering
\fbox{
\includegraphics[scale=.50]{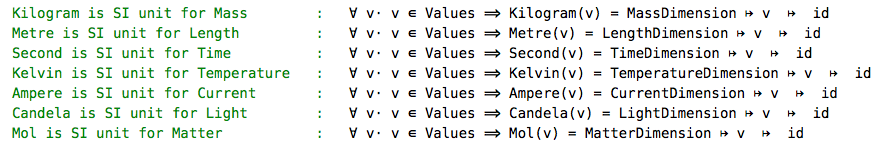}}
\caption{The 7 SI base unit measurements in {\tt CONTEXT C\_Measurements}}
\label{fig:BaseUnitMeasurements}  
\end{figure}

We define a context {\tt C\_Measurements} 
which {\tt EXTENDS} {\tt C\_Values} and {\tt C\_Dimensions} 
and combines them to define the standard SI units for the 7 base dimensions:
{\tt Kilogram,
Metre,
Second, Kelvin, Ampere, Candela}, and
{\tt Mol} (see Figure~\ref{fig:BaseUnitMeasurements}).
The key modelling decision was to model a measurement as a triple: the dimension, the value, and the
bijective normalization function for mapping the value to the standard canonical unit.
Thus, the actual unit of measurement is implicit in the function definition.
For example, a {\tt Mass} in {\tt Kilogram}s is constructed from a {\tt MassDimension}, its {\tt Value} and
the {\tt id} function (as it is already in the standard unit canonical form). Thus, {\tt Kilogram(v)} represents a 
mass of $v$ kg.

\begin{figure}[htb]
\centering
\fbox{
\includegraphics[scale=.38]{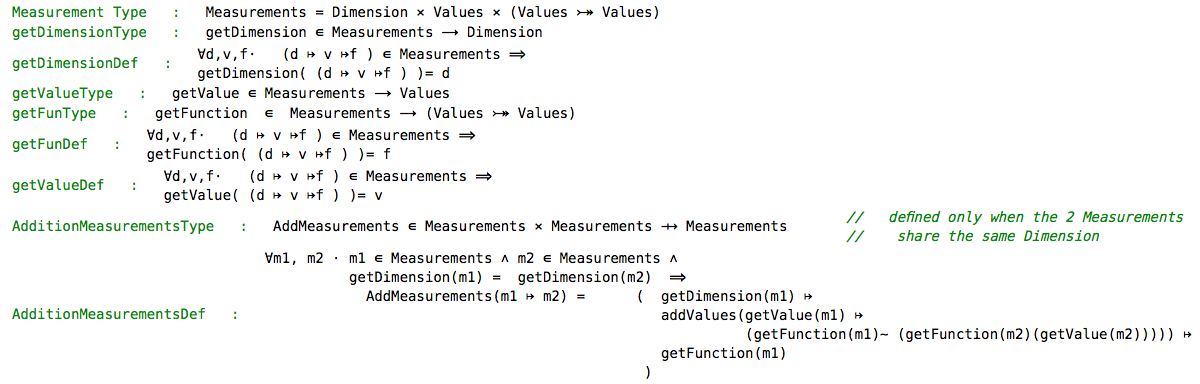}}
\caption{The addition of {\tt Measurements}}
\label{fig:ContextMeasurements-Addition}  
\end{figure}

We note that we make no assumption that units must be defined by a simple linear function through the origin
(as is the case for most physical scales). The measurement model also permits more general
affine functions (as for temperature conversion, for example).
The advantage of this explicit approach is that we can now formally specify the addition (and subtraction) of
{\tt Measurements} as partial functions (which are defined only when the {\tt Measurements} have the same 
{\tt Dimension}).
For simplicity our paper presents - in Figure~\ref{fig:ContextMeasurements-Addition} -
only the function for adding measurements. Subtraction is the same problem, whilst
multiplication and division require no consistency checking.
Similarly, we  model partial functions for comparing measurements, where the functions are defined (using the
comparison between the values converted to standard units) only when the measurements are in the same dimensions.

The key aspect of the mathematical operators is that the resulting {\tt Measurement} is given in the units of the
first {\tt Measurement} parameter. In order to do this, the second parameter {\tt Value} has to be converted to the 
same units. This is done by converting to the standard base unit of the {\tt Dimension} common to the two parameters, and then
converting to the unit of the first parameter. For example the addition of $100$ grammes and $2$ pounds, would result in
$(100+907,18474)$ grammes

\section{Illustrating measurement modelling  problems}
\label{sec:CaseStudyExamples}

In this section we review a small number of case studies which have included modelling of real-world
measures. We compare, with our explicit approach,  the way in which the studies (to various degrees) use implicit modelling
of units and dimensions.

\subsection{Changing the unit of measurement when formalizing a requirements specification}

In `{\it Modeling a Landing Gear System in Event-B}'~\cite{MammarLaleau14}, a refinement
is used to introduce timing aspects.
The authors note that with respect to time values: `{\it Since, the type Real is not provided in EventB, we multiply all the data per 10}'.
This impliclty suggests that the basic unit of time in their models will be one tenth of a second (also known as a decisecond).
This modelling decision is coherent with the original case study specification~\cite{BoniolWeils14}, where timing constraint values 
are mostly given in integer seconds, but timing measurement values (based on real world estimations) are given as floating point decimals (to 1 decimal place).
Both types of timing values are modelled as {\tt NAT}s in the Event-B models.
All the subsequent documentation that accompanies their model refers to units of time (u.t),
which are deciseconds, when  timing values (variables and constants) are modelled using the in-built {\tt NAT} set.

Even in this simple case, there is a problem of model validation and understandability.
The modellers must continually remember to multiply by ten all timing values given in the requirements specification.
Readers of the document who are trying to validate the model against their requirements must continually remember to
divide by ten all timing values given in the Event-B model. 
This type of mental gymnastics is illustrated by the requirement:
\begin{quote}
{\it If one of the three gears is not seen locked in the down position more than 10 seconds after stimulating the outgoing electro-valve, then the boolean output normal mode is set to false.}
\end{quote}
and
the corresponding Event-B condition:

\vspace*{-0.5cm}
\begin{center}
\includegraphics[scale=.50]{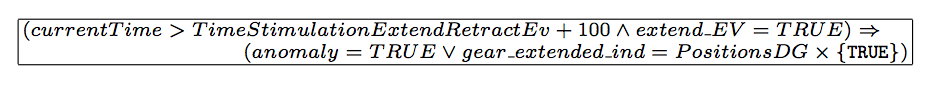}
\label{fig:LandingGearStudy1}  
\end{center}
\vspace*{-0.7cm}

A secondary concern with this modelling approach is maintainability.
If we imagine that a new requirement is added that uses a time measured to 2 decimal places,
it will now be necessary to globally change all time values to units of a hundredth of a second  (centiseconds).
This may not be as straightforward as just multiplying all integer {\tt NAT} values by 10, as {\tt NAT}s may also have been
used to model other variable types in the system (and not just times).
So we must be able to locate all time variable and constant definitions.
This may be done by introducing naming conventions (such as starting every time with the letter `T'); but such approaches are
notoriously brittle to changes to systems and models as they evolve. Further, such naming conventions
are not necessarily standard, and so composition and re-use of models which use them becomes non-trivial.

\begin{figure}[htb]
\centering
\fbox{
\includegraphics[scale=.45]{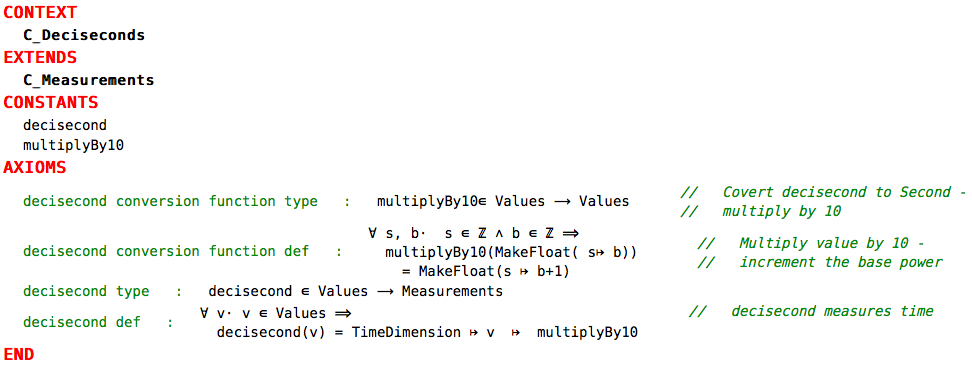}
}
\caption{The dimension model -  {\tt CONTEXT C\_Deciseconds}}
\label{fig:ContextDeciseconds}  
\end{figure}

Using our approach, we re-use the deciseconds measurement model
in the Event-B context shown  in Figure~\ref{fig:ContextDeciseconds}.

\vspace*{-0.3cm}

The expression, from the original model:\\
\small
\hspace*{12mm}
$currentTime > TimeSimulationExtendRetractEv+100$\\
\normalsize
 can then be replaced by
the expression:\\
\small
\hspace*{12mm}
$gt(currentTime \mapsto (AddMeasurements(TimeSimulationExtendRetractEv, Second(MakeFloat(1 \mapsto 1)) )$\\
\normalsize

The original model assumes that the variables $currentTime$ and $TimeSimulationExtendRetractEv$ are both times and that
they have the same units. Further, it assumes that the value $100$ is also a time given in the same units
as the two variables. With our explicit approach,
the fact that the two variables and one constant are all times will be checked by the Rodin tool as it verifies that the
expression is well-defined. Further, there is no need to statically check that the units are homogeneous; if they are
different then the conversions will be done implicitly. 
Finally,  it is easier to validate our model against the original natural language requirements
as the {\it 10} units of time specified in {\it seconds} 
are modelled explicitly as $1 \times 10^{1}$ {\tt Seconds}. (The {\it clumsy} syntax of our model will be improved in future work,
where we model measurements and values using a theory plug-in for the Rodin tool.)

\subsection{Overloading a {\it type/set/class} with values from a different scale of the same dimension}
\label{sec:NoseGearExample}  

In `{\it Integrating Domain-Based Features into Event-B: A Nose Gear Velocity
               Case Study'}~\cite{MeryST15}, we see a model which represents velocities using 3 different scales:
 Inch per millisecond (IPmS),
 Miles per hour (MPH), and
Kilometres per hour (KPH).
 There is an explicit typing of different base measurements: {\tt INCH},  {\tt MILE}, {\tt MILLISECOND}
 and {\tt HOUR}. However, 
 kilometres are not modelled explicitly as a type; but they are modelled implicitly in axioms {\it axm1} and {{\it axm2}:

 \begin{center}
\fbox{
\includegraphics[scale=.35]{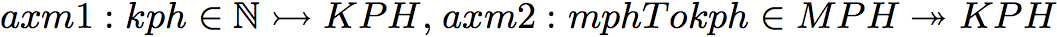}
}
\label{fig:ImplicitKILOMETERS}  
\end{center}
  
    We also see an explcit type for each of the velocities {\tt MPH} and {\tt KPH}, but
 no explicit type for {\tt IPmS}, which appears implicitly in axiom {\it axm6}.

  \begin{center}
\centering
\fbox{
\includegraphics[scale=.45]{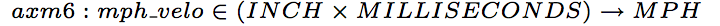}
}
\label{fig:INCHperMILLISECOND}  
\end{center}
 
 Axiom  {\it axm6} states only that  the velocity value is calculated from the inches value and the 
 milliseconds value, but there is no explicit statement that the length dimension is {\bf divided by} the time dimension.

 The explicit function (bijection) that maps {\tt MPH} to {\tt KPH} ({\tt mphTokph})
 implies that there is a dimensional equivalence (that of velocity), and we can therefore 
 coherently compare {\tt MPH} and {\tt KPH} values.
There is no explicit statement  that {\tt MILE}s and {\tt INCH}es represent the same dimension (length); and
that {\tt MILLISECOND}s and {\tt HOUR}s represent the same dimension (time).
The explicit types - as defined in their model -
will help us to formally express and validate the natural language requirements; but the 
lack of explicit dimensions can make the process complex.
For example, the requirement:
\begin{quote}
 ``{\it Estimated ground velocity of the aircraft is available only if it is within 
 3 KPH of the true velocity at some moment  within the past 3s''}
 \end{quote}
contains the constant value 3 for both a velocity and for a time.
The original specification of this behaviour\footnote{We have highlighted the constant values that
are linked to the natural language text.}  is given in Figure~\ref{fig:NoseGearStudy1}.

\begin{figure}[htb]
\centering
\includegraphics[scale=.38]{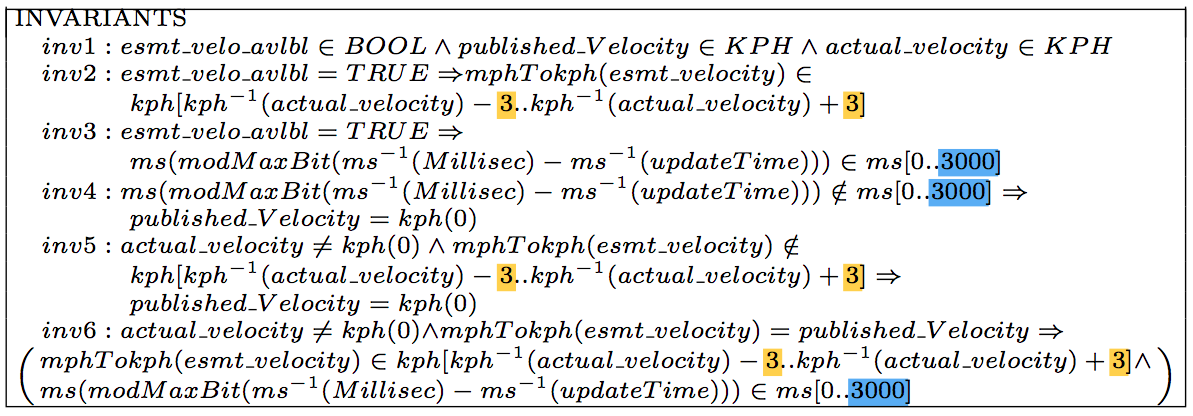}
\label{fig:NoseGearStudy1}  
\end{figure}

The ``3 KPH'' from the requirement text is  found in 6 locations - highlighted in yellow -
 in the formal model.
The ``3 s'' from the requirement text is  found in 3 locations - highlighted in blue -
 in the formal model.
Having an implicit rather than explicit typing of the values ``3'' in the model compromises understandibility and 
introduces complexity. As such, the model is more difficult to maintain and extend.

Following this modelling method, the coherence of arithmetic operations over values cannot be checked automatically.
For example, we have edited a sub-expression of
{\it inv2} in order to  mix velocity and time (written in red). The mismatch in types, in this case,  will not be  detected automatically
because the Rodin toolkit considers the expression to be well-defined.
 \begin{center}
\fbox{
\includegraphics[scale=.40]{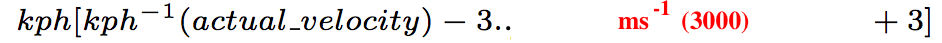}
}
\label{fig:IncoherentExpressions}  
\end{center}

In their approach,  all arithmetic is done over the {\tt NAT} values in the domain of the typing functions.
This facilitates the verification (proving) process, but hinders the validation process, and permits the 
modelling of mixed-type expressions that are ``meaningless'' in the real world.

This model also suffers from maintainability and scalability issues. Rather than explicitly modelling the
scaling of values between only non-canonical and canonical forms, they provide an explicit mapping between
2 non-canonical forms, namely {\tt MPH} and {\tt KPH}.
Such explicit mappings are good for
models with few units of measure, but they create scaleability problems when multiple units of measure are 
defined. For example, with 3 unit lengths - miles, kilometres, inches - and 2 unit times - hours and seconds - we
have 6 possible unit velocities. An explicit transformation function between each pair of unit velocities would
require 30 different function definitions in total. It would be much better to 
require only a single explicit function for mapping a value in each new single dimension unit to 
the canonical value in the same unit dimension; and for the transformations of all other units to be done in an automated, 
implicit, fashion. In this way, the model is much more scaleable to the introduction of new 
units.

Finally, the model also introduces ambiguous, non-standard notation. The model's invariants contain the expression  $ms^{-1}$.
In standard canonical form this would represent a velocity (metres per second); however, due to an unfortunate
choice of syntax, in their model it represents a mapping from a millisecond time to a {\tt NAT} value.
Such syntax clashes are inevitable when composing models, except if a standard formal ontology (as we propose)
is used by all developers.

With our approach we would explicitly model the new {\tt Velocity} dimension and the 3 different units using the
 {\tt AXIOM}s  as shown in Figure~\ref{fig:VelocityAxioms}  :
\begin{figure}[htb]
\centering
\fbox{
\includegraphics[scale=.42]{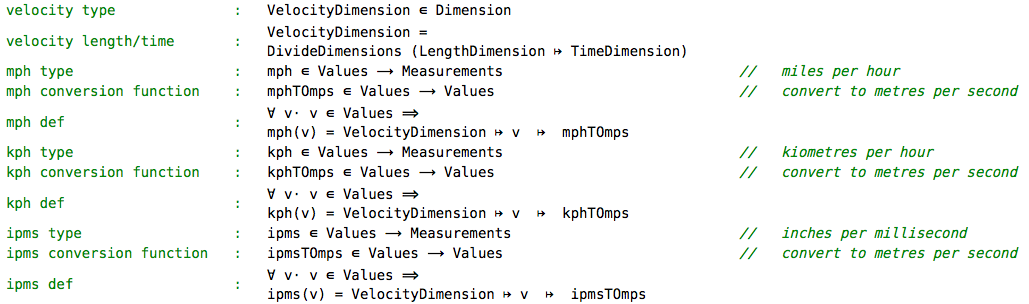}
}
\caption{The dimension model -  {\tt CONTEXT C\_Velocity  AXIOMS}}
\label{fig:VelocityAxioms}  
\end{figure}

Using the  {\tt Velocity}  context, we can now re-write {\it inv6} from the original model as:\\
\begin{quote}
$
(neq (actual\_velocity \mapsto kph(0))  \wedge eq(esmt\_velocity \mapsto published\_velocity)) \Rightarrow \\
( gt(esmt\_velocity \mapsto SubtractMeasurement(actual\_velocity \mapsto kph(3) \wedge \\
  lt(esmt\_velocity \mapsto AddMeasurements(actual\_velocity \mapsto kph(3)  )
$
\end{quote}

Coherency of the comparison functions ($neq$, $gt$ and $lt$) 
and the arithmetic functions ($SubtractMeasurement$ and $AddMeasurement$)
is guaranteed by the welldefinedness of the expression, as checked by the Rodin tool.

\subsection{Overloading a {\it type/set/class} with values from different dimensions}

In `{\it Formalizing hybrid systems with Event-B and the Rodin Platform}'~\cite{SuAbrialZhu14}, we see
the modellers overloading the positive reals to represent both temperature and time in a nuclear plant reactor cooling system:
\begin{quote}
{\it We define some state variables: $\theta$ is the temperature of the reactor, t1 and t2 denote the time elapsed since rod1 or rod2 have been released.}
\end{quote}
The modelling of the state of the real world measurements (temperature and time) and the discrete state of the rod is as follows:
 \begin{center}
\centering
\includegraphics[scale=.30]{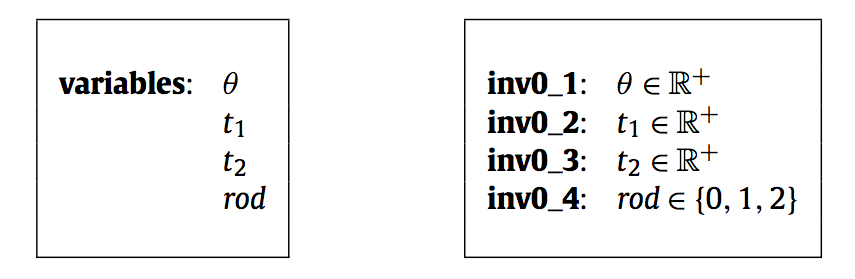}
\label{fig:Overloading}  
\end{center}

These variables are used in specifying the control events:
 \begin{center}
\centering
\includegraphics[scale=.35]{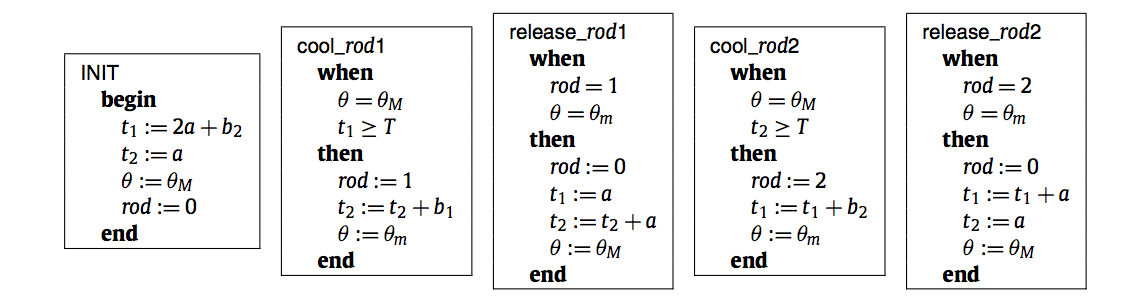}
\label{fig:EventsOverloadedTypes}  
\end{center}

With regard to the risk of adding values with different dimensions, we see expressions,  such as $2a+b_{2}$ and 
$t_{2}+b_{1}$, where
there is a risk that the operands are not coherent measures in the real world. The same risk is taken when comparing
values
(eg, $\theta = \theta_{M}$)
and assigning values
 (eg, $\theta := \theta_{m}$).
 Reading the model, it is clear that the authors have taken great care in naming their variables in order to help
 avoid any incoherent combination of the real values. However, the additional burden on the modellers to
 check for and guarantee coherence becomes ever more significant as the model  introduces implicit 
 composition of dimensions. A good example is the pump controller model, where a variable $v_{1}$ has implicit
dimension of length divided by time (a velocity):

\begin{figure}[htb]
\centering
\includegraphics[scale=.35]{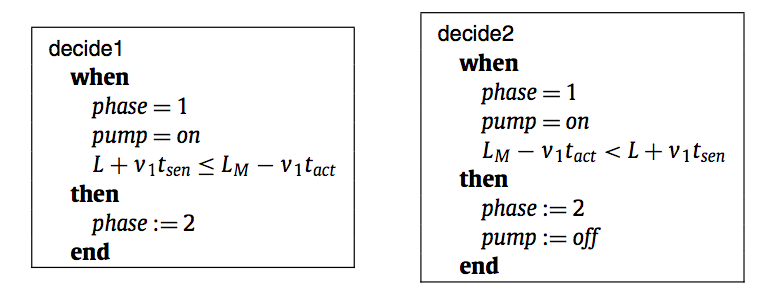}
\label{fig:ImplicitDimensionComposition}  
\end{figure}

In our approach, we already have explicit dimensions and measures for times and temperatures.
The only additional extra dimension required by their model is that of velocity; and so we can re-use the model
given in the previous section \ref{sec:NoseGearExample}.

\section{Application to other domains - currency example}
\label{sec:Currency}  

The previous sections have   demonstrated the utility of  our approach
with   respect to modelling   of the natural  physical world. However,
humans have also introduced abstract quantities into many domains, and
these  often  suffer from  the same   modelling   problems as  for the
concrete physical world.  A typical example is one of currencies. In a
{\it  shopping/services on-line domain},  it is often  possible to pay
using different  currencies (dollars,  euros, etc\ldots)  in different
subsystems. The  role of the  system would  be to  provide an ontology
matching when combining  such systems and domains~\cite{Euzenat07}.  We
need to be  able   to  check  that  we coherently   perform  numerical
operations using currency   values in the   same way  as for  physical
measurements.   This would  appear to  be a  trivial  extension to our
approach. However, it is made more complex by the conversion functions
between   currencies being variable    (rather  than constant, as  for
physical measurement conversions).  In order to model this in Event-B,
we are obliged to model the state of the system (in a machine).

In our example,  customers (in a set C)  and providers (in set  P) are
interacting, when  customers are requesting  services (in a set S). 
A service is provided by one and only one provider.
A  customer should first request    a service; the provider
sends him or her an  invoice and finally the  service  is paid by  the
customer. When the service is paid, it is delivered.

We have three levels of machines\footnote{The full models for this case study - and for the
previous measurement contexts - will be provided by the authors on request.} - \textsf{M0}, \textsf{M00} and \textsf{M000} -
corresponding to a progressive modelling of the problem as a sequence of refinements.

Machine  \textsf{M0}:  The customer is   requesting  a service  and  the
service is delivered by  the provider. No  expression of cost is given
and  no expression of failure in  transaction  is stated in our model.
Two events model the possible actions operating on variables deliver and order;  \textit{Ordering\_a\_service}   and \textit{Serving\_a\_requested\_service}. The invariant is simply stating that a  service is corresponding to a request ($deliver \subseteq order$).

Machine  \textsf{M00} introduces  the billing system in the process. A service should be  paid before being delivered. New variables related to billing are introduced, as well as the actions of   sending an invoice and paying an invoice. The invariant is simply stating the  sequentiality of different actions and the  strengthening of previous actions: 
\begin{center}
$\begin{Bcode}
{ inv1 }:{ pay \in  S\pfun C }\\
 {inv2 }:{billing \in  S\pfun C}\\
{ inv3 }:{ billing \subseteq  order }\\
{ inv4 }:{ pay \subseteq  billing }\\
{ inv5 }:{ deliver\subseteq pay }
\end{Bcode}$. 
\end{center}
\noindent
The machine has two new  events which are modelling the  two new variables: 	\textit{Billing\_a\_requested\_service} and 	\textit{Paying\_a\_requested\_service}. The event \textit{Serving\_a\_requested\_service}  has a new guard which is strengthening the  condition of service.

\begin{center}
$
\begin{Bcode}
		\quad { inv1 }:{ account \in  U\tfun Values }\\
		\quad { inv2 }:{ val \in  B\pfun Values }\\
		\quad { inv3 }:{ cust \in  B\pfun C }\\
		\quad { inv4 }:{ prov \in  B\pfun P }\\
		\quad { inv5 }:{ ser \in  B\pinj S }\\
		\quad { inv6 }:{ npay \in  B\pfun Values }\\
		\quad { inv7 }:{ date \in  B\pfun D }\\
		\quad { inv8 }:{ t \in  B\pfun (Values\tfun Values) }\\
		\quad { inv9 }:{ bills \subseteq  B }\\
\end{Bcode}\begin{Bcode}
		\quad { inv10 }:{ bills = dom(val) }\\
		\quad { inv11 }:{ bills=dom(cust) }\\
		\quad { inv12 }:{ bills=dom(prov) }\\
		\quad { inv13 }:{ bills=dom(ser) }\\
		\quad { inv14 }:{ tarif \in  S\tfun Values }\\
		\quad { inv15 }:{ dom(billing)=ran(ser) }\\
		\quad { inv16 }:{ dom(npay)\subseteq dom(ser) }\\
		\quad { inv17 }:{ dom(pay)=ser[dom(npay)] }\\
		\quad { inv19 }:{ dom(npay) \subseteq  bills }\\
\end{Bcode}
$
\end{center}

Machine \textsf{M000}  refines  machine  \textsf{M00} by \textit{superposing} new variables  in the current events. We introduce new variables such as  the variable  $bills$ which  models the  set of already  existing bills. The main  critical  property is the invariant 
$inv22$ which states that the value paid by the customer to the provider corresponds to the  value in the currency of the customer: $cpay(cust(b)\mapsto ser(b))=t(b)(val(b))$. $t(b)$ is  the  translation of the value of the  bill in the currency of the provider ($val(b)$)  and in the variables, we record the current translation used while paying.

\begin{center}
$
\begin{Bcode}
		\quad { inv20 }:{ \forall b\qdot  b \in  bills  \land  b \in  dom(npay) \limp  val(b)=npay(b) }\\
		\quad { inv21a }:{ cpay \in  C\cprod S\pfun Values \land  dom(npay)=dom(date)  }\\
		\quad { inv21b }:{  dom(t)=dom(date) \land  dom(npay)=dom(t)  }\\
		\quad { inv22 }:{ \forall b\qdot   
\left(\begin{array}{l}b \in  bills \land  b  \in  dom(date)\\ \limp\\   
cust(b)\mapsto ser(b) \in  dom(cpay) \land  cpay(cust(b)\mapsto ser(b))=t(b)(val(b))\end{array}\right. }\\
		\quad { inv23 }:{ dom(t)=dom(date) }\\
		\quad { inv24 }:{ dom(npay)=dom(t) }\\
\end{Bcode}
$
\end{center}

In this simple case study we see that the static approach for {\tt Measurements}, where the conversion functions
are constant, has the potential to be extended to a more dynamic approach (as for currency payments in the example)
where the conversion functions are variable.
It is currently work-in-progress to extend our methodology to handle the dynamic case.

\section{OO Design and Implementation }

\subsection{Inheritance Hierarchy}

We chose to implement our {\tt Measurements} model in Java because of the rich typing and 
polymorphic semantics, together with type safety for a reasonable subset of the language~\cite{Drossopoulou99}.
The diagram, in Figure~\ref{fig:MeasurementClassDiagram}, is a partial representation of the inheritance hierarchy in our object oriented implementation
(using Java).
There are seven single dimension measurements: Mass, Length, Time, Temperatures, Luminosity, Current and 
Substance, each of which has a standard unit class: Kilogram, Metre, Second, Kelvin, Candela, Ampere and Mole.
These are in the base library of Measurements that provide  the foundations for constructing three kinds of new classes:
\begin{itemize}
\item {\bf Non standard single dimensional measurements}, like Centimetre, Kilometre, Mile, Hour and Decisecond (in the diagram).
\item {\bf Standard multiple dimensional measurements}, like Velocity in MetresPerSecond  (in the diagram).
\item {\bf Non standard multiple dimensional measurements}, like Velocity in MilesPerHour (in the diagram).
\end{itemize}
Developers are free to add new Measurements (of all 3 types) by extending the class hierarchy. They must 
execute the Measurement unit tests (see section \ref{subsec:Unit tests}) on instances of their new classes in order to 
(partially) verify that the new classes respect the generic requirements as specified in the Event-B context models
(such as associativity and commutativity of the
addition operation/method). 

\begin{figure}[htb]
\centering
\fbox{
\includegraphics[scale=.45]{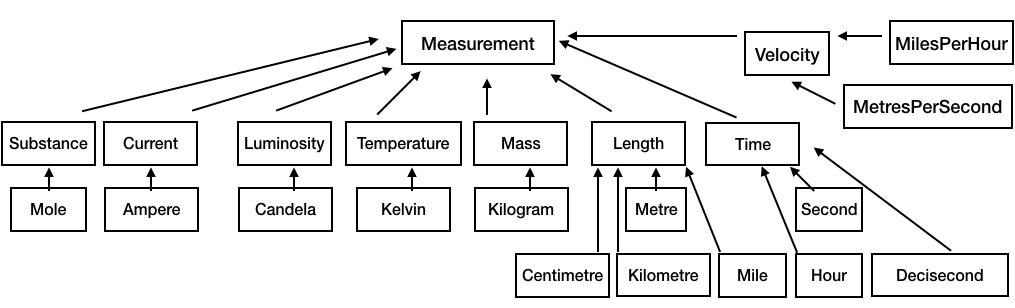}
}
\caption{(Incomplete) Measurement Class Inheritance Hierarchy}
\label{fig:MeasurementClassDiagram}
\end{figure}

\subsection{The Java Code}
The abstract behavior is first modelled in the class {\tt Measurement.java}, below.
Concrete measurement classes (in the standard seven dimensions and all possible combinations) are required to 
override the methods of this general class:
\begin{itemize}
\item the standard four numerical operations, with exceptions thrown for addition and subtraction of measurements
in different dimensions;
\item a method to permit the simple scaling of a measurement;
\item two methods for converting to/from the standard unit of measurement for the given dimension;
\item two methods (static and non-static) to check if  measures have the same dimension
\item two constructors (including a random constructor)
\item standard getter methods, toString, equals and hashcode
\end {itemize}
\begin{figure}[htb]
\centering
\fbox{
\includegraphics[scale=.55]{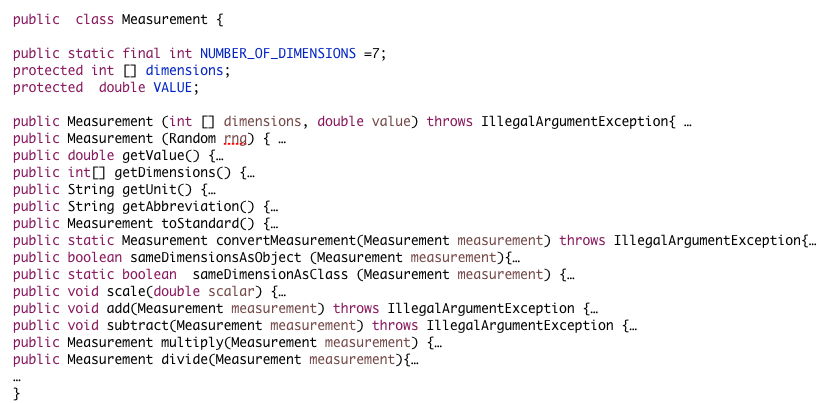}
}
\label{fig:Measurement-Java}
\end{figure}

The implementation of each base dimension measurement is now straightforward. For example, the Java code for the {\tt Second}
class is given, below. 

 \begin{center}
\centering
\fbox{
\includegraphics[scale=.55]{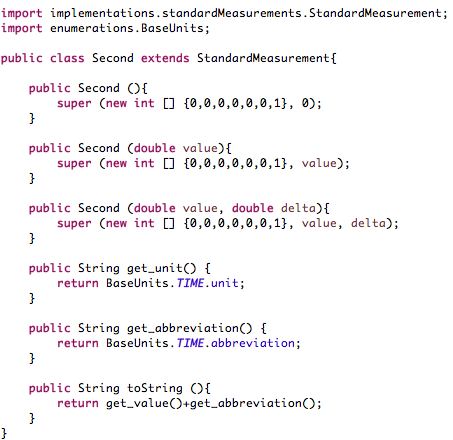}
}
\label{fig:Second-Java-Class}
\end{center}

The implementation of more complex single and multiple dimensional measurements is slightly more complicated
as it requires care when writing the methods for converting to/from the standard units.
We note that this is a simple scaling function (except in the case of temperatures).
The required properties (as specified in the Event-B theorems) are then used to drive the coding and
execution of unit tests on the new measurement classes.

\subsection{Java - unit tests from Event-B theorems}
\label{subsec:Unit tests}

Following the approach in ~\cite{GibsonLalletRaffy11II}, we use the formal models to drive the
generation of test cases, rather than refining them towards an implementation.
Thus, we identify theorems in the contexts (that have been proven by the Rodin tool) and
demonstrate that the properties that they specify are also valid for the Java implementation (through testing).
For example, the commutativity and transitivity of the addition of measurements (of the same dimension) can
be verified automatically by the RODIN tool, but needs to be tested in the Java implementation.

The technique that we employ is very simple, and based on random test generation combined with  the JUnit tool.
The abstract tests are coded in the JUnit_Measurement class, which must be extended by a unit test class
for each new Measurement class. For example,  the Second class has a test class JUnit\_Second which extends
 JUnit\_Measurement. The tests use  random constructors to check that the required properties
 are respected for a {\em large number of randomly generated} 
 test cases\footnote{We can run millions of tests in a matter of seconds.}
 This does not guarantee that the properties are respected for {\em all} the new measurement values, but they reassure the
 developers that there is a very high probability that their code is correct (with respect to the properties being tested).

\section{Ongoing and Future Work}

We are currently pursuing three areas of research which extend this work:

\noindent {\bf I. Accuracy of measurement -}
In  this paper we have  not addressed the  need to model the fact that
values and measurements in our systems  are actually approximations to
reals.  In certain types of  (relatively simple) systems this can lead
to               unpredictability            and               chaotic
behaviour~\cite{Gleick11}.  Consequently, in order  to reason about the
behaviour of such systems,   it is necessary   for values to  modelled
within a certain error (within a bounded interval). Fortunately, there
is  already   much  existing  work  on interval    values and interval
arithmetic
~\cite{Moore62, Kearfott96,Kearfott96,HickeyJV01,BronnimannMP03}.
We have already extended our Java measurement library to include interval value arithmetic.
We hope to report on this in a future paper.

\noindent {\bf II. Theory of measurements -}
Rodin provides a {\it Theory plug-in}  that facilitates definitions of  mathematical and prover
 extensions~\cite{HoangVSBWB17}.
It should be relatively straightforward to introduce a theory 
of {\tt Measurements} using the current algebraic specifications in our {\tt Values}, {\tt Dimensions}
and {\tt Measurements} contexts. However, it seems wise to wait until our work
on modelling values as intervals is complete and validated before we attempt to build a theory extension
for interval measurements.
It should be noted that such a theory extension would help to tidy up our current {\it clumsy} notation, and
would facilitate more automated proof of simple mathematical properties of measurements.

\noindent {\bf III. Re-engineering case studies for validation and extension of our approach -}
It is important to validate our approach using real world case studies. We are currently re-engineering a number of case studies
that have used measurements implicitly to adopt our explicit measurement modelling approach.
We believe that these re-engineered models - thanks to the explicit modelling of measurements - will be easier
to validate, more amenable to automated coherency checking, and easier to maintain.
One of the case studies that we are currently re-engineering is the currency example - mentioned previously in
this paper -  where there is a need to model and reason about dynamic rather than static unit conversion.

\section{Conclusions}

We have motivated the need for explicit modelling of dimensions of physical measurements during 
formal development of systems. We have demonstrated one approach for such modelling using Event-B contexts, that
meets all our formal engineering requirements.
We have compared this approach with other case studies in which the modelling of dimensions is implicit.
Java implementations of the Event-B specifications demonstrate the feasibility of the approach.
Java unit tests are then derived from the Event-B specifications.

\section*{Acknowledgement}

The authors acknowledge the support provided by the French ANR project IMPEX (13-INSE-0001)

\bibliographystyle{eptcs}
\bibliography{IMPEX2017}
\newpage

\end{document}